\documentclass[12pt]{iopart}
\usepackage[pdftex]{graphicx}
\usepackage{color}
\usepackage{hyperref}
\usepackage{siunitx}
\usepackage[superscript]{cite} 
\usepackage{iopams}  
\begin{document}

\title[Voltage-dependent first-principles simulation of chloride insertion]{Voltage-dependent first-principles simulation of insertion of chloride ions into Al/Al$_2$O$_3$ interfaces using the Quantum Continuum Approximation}

\author{Quinn Campbell}
\address{Sandia National Laboratories, Albuquerque, NM USA}
\ead{qcampbe@sandia.gov}
\vspace{10pt}
\begin{indented}
\item[]\today
\end{indented}

\begin{abstract}
Experiments have shown that pitting corrosion can develop in aluminum surfaces at potentials $> -0.5$ V relative to the standard hydrogen electrode (SHE).
Until recently, the onset of pitting corrosion in aluminum has not been rigorously explored at an atomistic scale because of the difficulty of incorporating a voltage into density functional theory (DFT) calculations. 
We introduce the Quantum Continuum Approximation (QCA) which self-consistently couples explicit DFT calculations of the metal-insulator and insulator-solution interfaces to continuum Poisson-Boltzmann electrostatic distributions describing the bulk of the insulating region.
By decreasing the number of atoms necessary to explicitly simulate with DFT by an order of magnitude, QCA makes the first-principles prediction of the voltage of realistic electrochemical interfaces feasible.
After developing this technique, we apply QCA to predict the formation energy of chloride atoms inserting into oxygen vacancies in Al(111)/$\alpha$-Al$_2$O$_3$ (0001) interfaces as a function of applied voltage. 
We predict that chloride insertion is only favorable in systems with a grain boundary in the Al$_2$O$_3$ for voltages $> -0.2$ V (SHE).
Our results roughly agree with the experimentally demonstrated onset of corrosion, demonstrating QCA's utility in modeling realistic electrochemical systems at reasonable computational cost. 
\end{abstract}

%
%
%
%
%

\section*{Introduction}

Corrosion of metallic interfaces leads to weakening of their structural integrity and has been widely estimated to cost billions of dollars annually.\cite{koch2002corrosion}
While this has inspired numerous studies on corrosion and strategies to protect against it, the fundamental atomistic mechanisms behind corrosion are difficult to identify for any given system. 
Corrosion in aluminum provides an excellent example of the issues.
In aluminum, it has been experimentally demonstrated that pitting corrosion begins at potentials $\Phi > -0.50$ V relative to the standard hydrogen electrode (SHE), dependent on the salt and pH concentration of the surrounding solution.\cite{dibari1971electrochemical,bessone1992eis}
The exact mechanism that causes this pitting corrosion is still under debate, however. 
Pitting corrosion is often described using the point defect model,\cite{lin1981point,chao1981point,engelhardt2004unification} which states that corrosion is initially driven by the formation of charged defects in the surface of the material, of which chloride ions are particularly damaging.\cite{frankel1998pitting,natishan20182017} 
On aluminum surfaces, a several nm thick aluminum oxide layer forms \cite{evertsson2015thickness} and the insertion of chloride into oxygen vacancies in this layer has been theorized as one of the main mechanisms leading to corrosion. 
Experiments have investigated the dependence of pitting corrosion on chlorine concentration in the surrounding electrolyte and other factors,\cite{brown1973use,shimizu1999novel,serna2006critical} but the results have not been fully conclusive as to the causal mechanism.

While many previous density functional theory (DFT) studies have examined Al/Al$_2$O$_3$ interfaces at an atomic level,\cite{kim2013nature,batyrev2001plane,koberidze2018structural,costa2014atomistic,siegel2002adhesion} they have typically not included oxygen vacancies and chloride insertion. 
For those that have examined vacancies,\cite{weber2009point,carrasco2004theoretical,hine2009supercell,janetzko2004first} they do not include Al metal in the calculation. 
Previous DFT studies on chloride insertion into Al/Al$_2$O$_3$ interfaces that have included contributions from the metal, the oxide, vacancies, and chloride ions \cite{liu2019dft,liu2021density,leung2021first} have either implicitly limited their scope to defects with no charge, or tested only a small range of voltages.

This lack of study of charged ions in the full electrochemical environment stems from the difficulty of needing to include several different length scales in a simulation to get a realistic picture of corrosion behavior.
Modeling the realistic interaction of point defects with the surface requires a quantum mechanical treatment, with DFT providing a convenient compromise between accuracy and speed.
DFT is typically limited, however, to simulating $\sim$1000s of atoms as larger systems are computationally prohibitive. 
Since a realistic model of an interface, including the surrounding solution, can be 10s to 100s of nm wide, DFT simulations must necessarily focus on a single region of interest, typically the surface. 
The entire surrounding electrode helps determine the electric field (and therefore voltage) that the region of interest operates at.
Simulations of these isolated regions therefore do not provide information as to the voltage the electrode is under and, in effect, operate at ``flatband'' conditions where the electric field is flat across the system. 
This necessarily limits the scope of traditional DFT calculations when examining these interfaces and defects. 
Since the defects under examination are charged, it is crucial to incorporate a realistic model of voltage into DFT calculations. 

\begin{figure*}
    \centering
    \includegraphics[width=\textwidth]{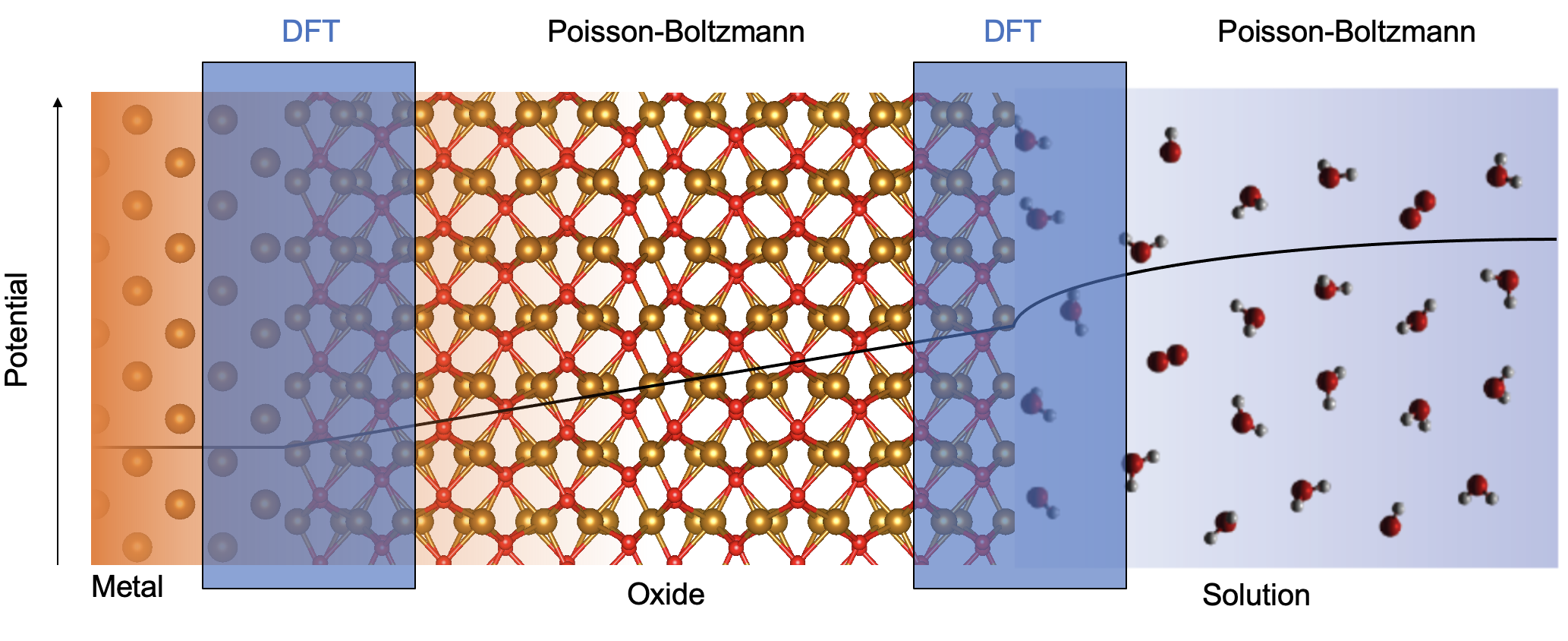}
    \caption{A schematic illustration of the Quantum Continuum Approximation applied to a metal-insulator interface immersed in solution.}
    \label{fig:qca-schematic}
\end{figure*}

In this paper, we develop a Quantum Continuum Approximation (QCA) methodology that allows for the first-principles simulation of the metal-insulator interfaces at controlled voltages using limited DFT simulation cell sizes.
QCA self-consistently couples DFT calculations of the metal-insulator interface and the insulator-solution interface with Poisson-Boltzmann distribution of charges in the bulk insulator and solution to provide an equilibrium potential drop for the entire electrode, as schematically illustrated in Fig.~\ref{fig:qca-schematic}. 
By reducing the number of atoms needed for DFT calculations to include voltage information by at least an order of magnitude, QCA enables voltage-dependent first-principles simulation of realistic electrochemical interfaces at reasonable computational cost.
We demonstrate this methodology for the simulation of chloride insertion into the aluminum oxide layer of an Al(111)/$\alpha$-Al$_2$O$_3$(0001) system as a function of voltage. 
We predict that insertion is not favorable for Al/Al$_2$O$_3$ interfaces that do not include grain boundaries at any voltage. 
We do see, however, that in systems with grain boundaries in the Al$_2$O$_3$, chloride insertion is favorable at $\Phi > -0.2$ V (SHE), in reasonable agreement with experiment and helping confirm the importance of grain boundaries in the corrosion process. 
This work provides a template for attaching a potentiostat to electrochemical DFT calculations of interfaces.

\section*{Methods}
\subsection*{Chloride insertion energy calculations}

We calculate the formation energy of a Cl$^-$ ion inserting into an oxygen vacancy as  
\begin{equation}
    \Delta E_{\text{Cl-O}}[\Phi] = E_{\text{DFT}}(\text{Cl ins.})[\Phi] - E_{\text{DFT}}(\text{O vac.})[\Phi] - E_{\text{Cl}^-}[\Phi] ,
    \label{eq:total-e}
\end{equation}
where $E_{\text{DFT}}(\text{O vac.})[\Phi]$ is the DFT energy of the slab with the oxygen vacancy as a function of the global potential $\Phi$ of the system, $E_{\text{DFT}}(\text{Cl ins.})[\Phi]$ is the DFT energy of the slab with a chlorine ion inserted into the previous location of the oxygen vacancy as a function of the global potential, and $E_{\text{Cl}^-}[\Phi]$ is the energy of the Cl$^-$ ion in solution as a function of voltage.

For the chloride ion, we follow a computational hydrogen electrode-like approach,\cite{norskov2004origin,rossmeisl2007electrolysis,man2011universality} where we determine the energy of the chloride ion as 
\begin{equation}
    E_{\text{Cl}^-}[\Phi] = \mu (\text{Cl}^-) + e\Phi^{\text{SHE}}.
    \label{eq:chloride-e}
\end{equation}
Here, $\mu (\text{Cl}^-)$ is the equilibrium chemical potential of the chloride ion, and $\Phi^{\text{SHE}}$ is the potential that the electrode is operating at, referenced to the standard hydrogen electrode (SHE).
We determine the chemical potential by referencing the following chemical reaction
\begin{equation}
    \text{Cl}_2 (\text{g}) + 2 e^- \rightarrow 2\text{Cl}^-(\text{aq}),
\end{equation}
which is at equilibrium at $\Phi_{\text{Cl}_2/\text{Cl}^-} = 1.36$ V  (SHE).\cite{atkins2006physical}
Using this reaction, we can then calculate the chemical potential of a chloride ion as 
\begin{equation}
    \mu (\text{Cl}^-) =\frac{1}{2} E_{\text{DFT}}(\text{Cl}_2(\text{g})) - e\Phi_{\text{Cl}_2/\text{Cl}^-}.
\end{equation}
This energy calculation is inherently dependent on the calculated dissociation energy of the Cl$_2$ molecule, which we calculate as 3.46 eV with PBE, compared to a value of 2.47 eV from experiment.\cite{le1971dissociation}
However, to maintain consistency with the PBE calculations of Cl insertion throughout the work, we have not applied any corrections to the calculated PBE energies as this would interfere with the same overbinding tendency when using PBE to calculate the energy of inserting Cl into a slab. 
With this energetic relationship established, we can now calculate the formation energy of chloride insertion into the Al/Al$_2$O$_3$ surface as a function of the electrode potential.
For this relationship to be strictly accurate, however, the calculated DFT energies of the slabs need to be taken at simulation conditions consistent with $\Phi$ for the electrode.
We outline how we can achieve this using the QCA methodology in the next section.

\subsection*{Quantum Continuum Approximation (QCA) Methodology for Predicting Voltage at an interface}
\label{subsec:qca}
\subsubsection*{Interface Energetics and Electrostatics}
At a fundamental level, our approach to calculations with metal-semiconductor-solution interfaces works by assigning a global potential to the system. 
This potential drop can then be split up between several different voltage drop regions within the interface.
Finding the correct distribution of potential drops requires a self-consistent process for minimizing the free energy of the system while taking into account the electrostatic requirements of following the Poisson equation and maintaining electronic equilibrium throughout the interface. 

To accomplish this task, we employ a new extension of the Quantum Continuum Approximation (QCA) methodology that was first developed by Campbell and Dabo for semiconductor-solution interfaces,\cite{campbell2017quantum,campbell2019voltage} which allowed for the voltage of the system to be additionally predicted by DFT slab calculations. 
It was later extended to solid-state metal-semiconductor interfaces, providing predictions of Schottky barriers that corresponded well with nano-ARPES measurements.\cite{subramanian2020photophysics} 
Here, we extend this to include metal-insulator or metal-semiconductor interfaces that are also immersed in solution, which are common in electrochemical applications. 
The basic goal is to self-consistently couple DFT calculations of an interface with analytic Poisson-Boltzmann descriptions of the bulk semiconductor and solution regions.
As shown in Fig.~\ref{fig:qca-schematic}, we perform DFT calculations only on the crucial interface sections, such as the metal-metal oxide transition and the metal oxide-solution transition, or around atomic defects in the system.
We then apply Poisson-Boltzmann calculations to the oxide  regions in between the DFT calculations and the surrounding solution. 

The free energy of the system is 
\begin{equation}
    \mathcal{F} = \mathcal{F}^{\text{m}|\text{ox.}} + \mathcal{F}^{\text{ox.}} + \mathcal{F}^{\text{ox.}|\text{sol.}} + \mathcal{F}^{\text{sol.}} - \frac{1}{2}\int dr \epsilon_0 \epsilon(\mathbf{r}) |\nabla(\phi(\mathbf{r}))|^2 ,
\end{equation}
where $\epsilon_0$ is the vacuum permittivity, $\epsilon(\mathbf{r})$ is the dielectric constant as a function of position, $\phi$ is the electrostatic potential, and $\mathcal{F}^{\text{m}|\text{ox.}}$ and $\mathcal{F}^{\text{ox.}|\text{sol.}}$ are the free energies of the metal-oxide and oxide-solution interface, respectively, which can be calculated using the typical calculations for DFT energy. 
Similarly, $\mathcal{F}^{\text{ox.}}$ and $\mathcal{F}^{\text{sol.}}$ are the contributions of the bulk oxide and solution regions, which can be calculated as analytic functions of the distribution of charge carriers within each system.
This decomposition of the free energy naturally leads to the following Poisson equation for the system:
\begin{equation}
    \nabla(\epsilon_0\epsilon(\mathbf{r})\nabla \phi(\mathbf{r})) = p(\mathbf{r}) - n(\mathbf{r}) -p_d(\mathbf{r}) + n_d(\mathbf{r}) + c_+(\mathbf{r}) - c_-(\mathbf{r}) + \rho_+(\mathbf{r}) -\rho_-(\mathbf{r}),
    \label{eq:electrostatics}
\end{equation}
where and $\rho_-(\mathbf{r})$ is the electron density obtained from the self-consistent Kohn-Sham equation for a given distribution $\rho_+(\mathbf{r})$ of atomic cores, and the source terms can be expressed as 
\begin{equation}
 n(\mathbf{r}) = n_d(\mathbf{r}) \left[1 + \exp\left(\frac{\phi(\mathbf{r}) + \epsilon_c - \epsilon_F}{k_{\text{B}}T}\right)\right]^{-1},
\end{equation}
\begin{equation}
 p(\mathbf{r}) = p_d(\mathbf{r}) \left[1 + \exp\left(\frac{\epsilon_F-\epsilon_v-\phi(\mathbf{r})}{k_{\text{B}}T}\right)\right]^{-1},
\end{equation}
\begin{equation}
 c_{\pm}(\mathbf{r}) = c^{\circ}(\mathbf{r}) \exp\left(\mp \frac{\phi(\mathbf{r})}{k_{\text{B}}T}\right).
\end{equation}
Here, $\epsilon_F$ denotes the Fermi energy, $\epsilon_v$ is the energy at the valence band maximum, $\epsilon_c$ is the energy at the conduction band minimum, $n_d(\mathbf{r})$ is the concentration of $n$-type carriers in the semiconducting/insulating region and is 0 outside this region, $p_d(\mathbf{r})$ is similarly the concentration of $p$-type carriers within the semiconducting regions, and $c^{\circ}(\mathbf{r})$ is the equilibrium ionic concentration inside the electrolyte region and is 0 outside of the electrolyte region.
In an $n$-type semiconducting region, the charge carriers can be solved for using the Boltzmann distribution as
\begin{equation}
    p(\mathbf{r}) = p_d = 0,
\end{equation}
\begin{equation}
    n(\mathbf{r}) = n_d\exp\left(\frac{\phi_0 - \phi(\mathbf{r})}{k_{\text{B}}T}\right),
\end{equation}
where $\phi_0$ is the asymptotic value of the potential in this region. 
Analogous equations can be written for $p$-type semiconducting regions.

\subsubsection*{Process for solving electrostatics}

To practically solve Eq.~\ref{eq:electrostatics} for a given potential, we start by dividing the system into two portions that we will simulate with DFT separately. 
The first is the surface of the interface, i.e. where the passivating oxide layer meets the solution.
We will refer to this region with a superscript ox.$|$sol. in variables throughout this work.
The other DFT region is the interface of the metal with the passivating oxide, referred to as a superscript m$|$ox. within this work. 
For both of these regions we start by doing DFT calculations on these systems when no charge has been applied.
This corresponds to the so-called `flatband' conditions. 
We then measure both the planar averaged potential profile of these systems $\phi_{\text{fb}}^{\text{m}|\text{ox.}}(z)$ and $\phi_{\text{fb}}^{\text{ox.}|\text{sol.}}(z)$, and the Fermi level of these systems $\varepsilon_{\text{F,fb}}^{\text{m}|\text{ox.}}$ and $\varepsilon_{\text{F,fb}}^{\text{ox.}|\text{sol.}}$.

Next we assign a total charge $Q$ to the electrode. 
The electrode is here defined as the region from the metal to the interface with the solution. 
The distribution of this charge $Q$ within the interface mechanically determines the total potential drop via Gauss' law.
For a perfectly insulating/semiconducting system, the charge distribution can be well approximated by the following Poisson--Boltzmann relationship:\cite{schmickler2010interfacial}
\begin{equation}
    \frac{d^2\phi}{dz^2} = \frac{n_{\text{d}}}{\epsilon_o \epsilon_{\text{sc}}}  \left[ 1 - \exp \left( \frac{-\phi}{k_{\text{B}}T}\right)\right].
    \label{eq:pb-ms-conditions}
\end{equation}
Here, $n_{\text{d}}$ is the concentration of charge carriers (dopants) within the semiconductor, $\epsilon_o$ is the vacuum permittivity, $\epsilon_{\text{sc}}$ is the dielectric constant of the semiconductor, and $k_{\text{B}}$ is the Boltzmann constant. 

For most real-world systems, however, surface states will form at both the metal-insulator and insulator-solution interface.\cite{bardeen1947surface,cowley1965surface,bard1980role} 
These surface states result from the abrupt nature of the interface disrupting the symmetry of the material, and from defects induced by interaction. 
These surface states trap charge at the interface, thus reducing the charge throughout the remainder of the insulator and typically reducing the potential drop of the electrode.
Capturing these surface states requires a quantum mechanical treatment of the interfaces using, e.g., DFT.

\begin{figure*}
    \centering
    \includegraphics[width=\textwidth]{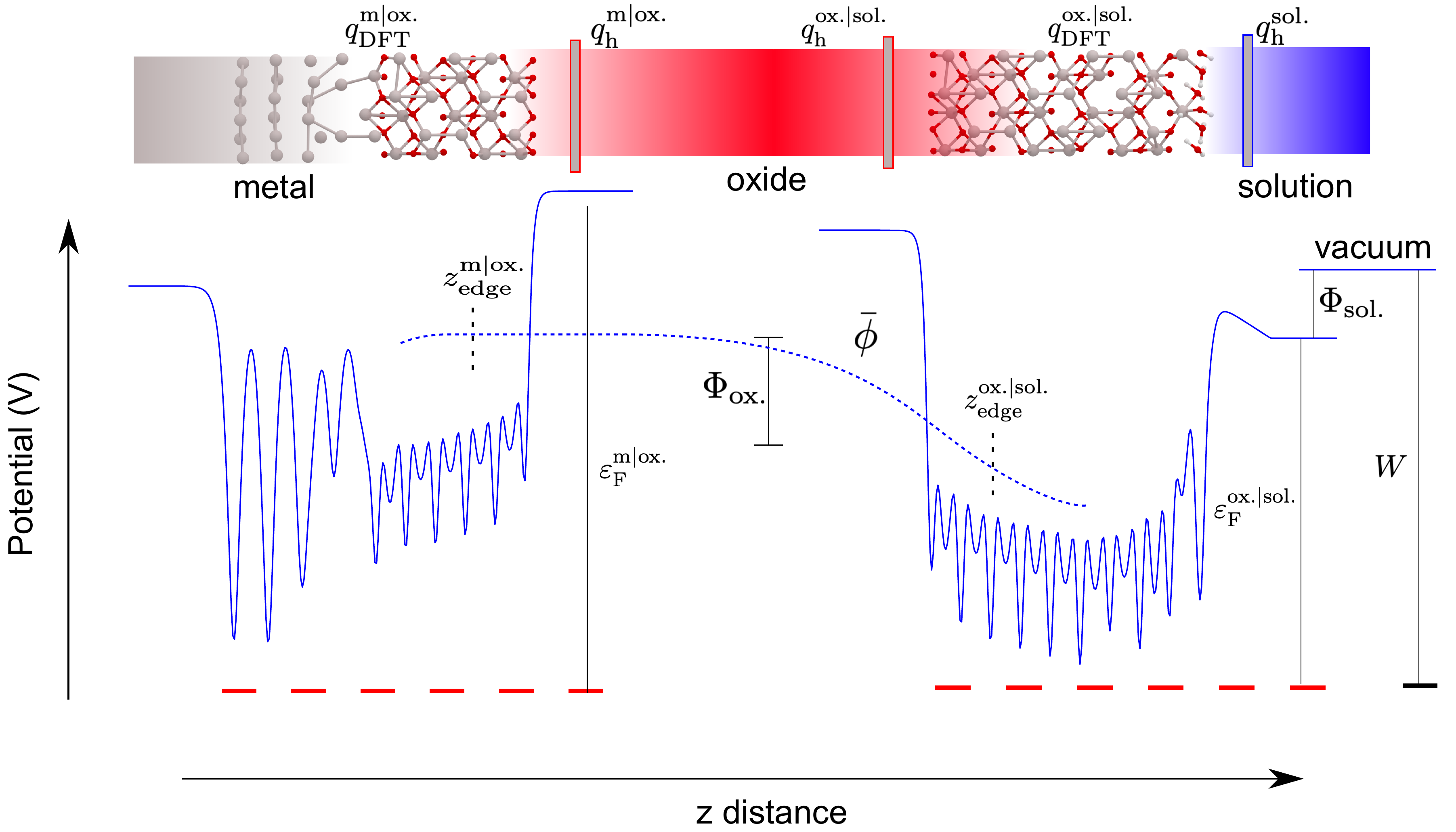}
    \caption{A schematic illustration of the potential profile of a metal-oxide-solution interface within QCA with important variables depicted. Distances and potentials shown are approximate.}
    \label{fig:qca-potentials}
\end{figure*}

To simulate charge accumulation at these surface interfaces we place Helmholtz planar counter-charges on either side of a DFT slab, as illustrated in Fig.~\ref{fig:qca-potentials}.
We place the Helmholtz planes 4 \AA\ away from the surfaces of the slab to mimic the Debye length, which is estimated between 3-5 \AA\ for typical ion concentrations in water.
Previous work has found the exact distance of the Helmholtz plane in this range to not significantly affect results.\cite{keilbart2017quantum} 
We obey charge neutrality for the DFT systems simulated, such that
\begin{equation}
    q_{\text{h}}^{\text{ox.}|\text{sol.}} + q_{\text{DFT}}^{\text{ox.}|\text{sol.}} + q_{\text{h}}^{\text{sol.}} = 0 
    \label{eq:chg-neutrality-ox-sol}
\end{equation}
at the oxide-solution interface, where $q_{\text{h}}^{\text{ox.}|\text{sol.}}$ and $q_{\text{h}}^{\text{sol.}}$ is the charge on the Helmholtz planes on the oxide and solution side of the DFT slab, respectively. 
Additionally, $q_{\text{DFT}}^{\text{ox.}|\text{sol.}}$ is the charge on the DFT slab representing the interface between the oxide and solution.
Since the electrode charge $Q$ is defined as including the metal-oxide interface, the continuum oxide, and the oxide-solution interface, we can assign the charge on the Helmholtz plane representing the solution as the opposite of this electrode charge, \textit{i.e.} $q_{\text{h}}^{\text{sol.}} = -Q$, maintaining charge neutrality within our larger system.

This arrangement of charge creates a potential gradient across the system.
We measure this as $\bar{\phi}^{\text{ox.}|\text{sol.}}(z) = \phi^{\text{ox.}|\text{sol.}}_{q} - \phi^{\text{ox.}|\text{sol.}}_{\text{fb}}$, where $\phi^{\text{ox.}|\text{sol.}}_{q}$ is the planar averaged potential of the charged system and $\bar{\phi}^{\text{ox.}|\text{sol.}}(z)$ is the difference in the planar averaged potential due to the charge. 
We can finally couple this DFT potential to a bulk charge distribution by extracting the derivative of the subtracted potential $\frac{d\bar{\phi}}{dz}$ at a specific cutoff point $z_{\text{edge}}^{\text{ox.}|\text{sol.}}$.
In DFT, our oxide will necessarily only extend for a few layers, but in reality it should extend for at least several nanometers.
This nonphysical oxide surface creates spurious surface effects on the potential. 
By setting the cutoff value for the potential $z_{\text{edge}}^{\text{ox.}|\text{sol.}}$ a few layers within the material, we avoid these surface impacts changing the potential away from the correct bulk-like potential. 

We can now use this information to inform the potential distribution of the bulk oxide region using the Poisson-Boltzmann conditions outlined in Eq.~\ref{eq:pb-ms-conditions}. 
The potential within the continuum oxide region can then be approximated as:
\begin{equation}
    \bar{\phi}^{\text{ox.}}(z) = \bar{\phi}(z_{\text{edge}}^{\text{ox.}|\text{sol.}}) + \frac{\epsilon_o \epsilon_{\text{sc}}}{2n_{\text{d}}}\left[\left(\frac{d\bar{\phi}^{\text{ox.}}}{dz}(z)\right)^2 - \left(\frac{d\bar{\phi}}{dz}(z_{\text{edge}}^{\text{ox.}|\text{sol.}})\right)^2\right] - k_{\text{B}}T .
    \label{eq:ox-pot}
\end{equation}
This relationship applies throughout the bulk oxide region, which lasts from $z_{\text{edge}}^{\text{ox.}|\text{sol.}}$ to $z_{\text{edge}}^{\text{m}|\text{ox.}}$.
We will specify the potential drop across this region as $\Phi_{\text{ox.}} = \bar{\phi}(z_{\text{edge}}^{\text{m}|\text{ox.}}) -\bar{\phi}(z_{\text{edge}}^{\text{ox.}|\text{sol.}}) $.

Similarly, for the metal-oxide interface we can say that 
\begin{equation}
    q_{\text{DFT}}^{\text{m}|\text{ox.}} +  q_{\text{h}}^{\text{m}|\text{ox.}}= 0 ,
    \label{eq:chg-neutrality-m-ox}
\end{equation}
where $q_{\text{h}}^{\text{m}|\text{ox.}}$ is the charge on the Helmholtz planes on the oxide sides of the DFT slab, and $q_{\text{DFT}}^{\text{m}|\text{ox.}}$ is the charge on the DFT metal-oxide interface.
We note that for metals, all excess charge can be expected to accumulate at the interface, leading to a flat potential within the bulk metal. 
This means that it is unnecessary to include a symmetric Helmholtz plane representing charge within the rest of the metal electrode.
Here, the value of $q_{\text{h}}^{\text{m}|\text{ox.}}$ should be set equal to the amount of charge that has accumulated in the system up to that point. 
Using Gauss' law, this would mean that 
\begin{equation}
q_{\text{h}}^{\text{m}|\text{ox.}} = \epsilon_o \epsilon_{\text{sc}} A \frac{d\bar{\phi}^{\text{ox.}}}{dz}\left(z_{\text{edge}}^{\text{m}|\text{ox.}}\right),
\label{eq:int2-chg}
\end{equation}
where $A$ is the lateral surface area of the metal-oxide DFT interface. 
Since this charge will be localized to the metal-oxide oxide interface in the subsequent DFT calculation, the area of the metal-oxide DFT interface is therefore the relevant one for Gauss' law.

While this setup ensures continuity between the two different DFT equations, it does not yet provide guidance on what the correct distribution of charges in Eq.~\ref{eq:chg-neutrality-ox-sol} should be.
The final condition to find the correct distribution of charges is to specify that a system at equilibrium must have the Fermi level be the same throughout the system:
\begin{equation}
    \varepsilon_{\text{F}}^{\text{ox.}|\text{sol.}} = \varepsilon_{\text{F}}^{\text{m}|\text{ox.}} + \Phi_{\text{ox.}},
\end{equation}
where $\varepsilon_{\text{F}}^{\text{ox.}|\text{sol.}}$ and $\varepsilon_{\text{F}}^{\text{m}|\text{ox.}}$ are the Fermi level of the DFT oxide-solution interface and DFT metal-oxide interface, respectively.
By adding in the potential shift from the oxide $\Phi_{\text{ox.}}$, we make sure these are evaluated from the same reference point.
We can relate these DFT calculated Fermi levels to the vacuum energy referenced work function by adding on a potential shift from the solution $\Phi_{\text{sol.}}$:
\begin{equation}
    W = \Phi_{\text{sol.}} + \varepsilon_{\text{F}}^{\text{ox.}|\text{sol.}} .
\end{equation}
This solution potential shift can be determined directly by another Poisson-Boltzmann shift,\cite{fisicaro2016generalized,mathew2019implicit} or by a constant solvation shift.
In real world scenarios, this potential drop would further be determined by the salt concentration and pH of the solution.
These effects can be partially included when calculating the formation energy of ions as in Eqn.~\ref{eq:chloride-e}, and can also be somewhat included in the implicit Poisson-Boltzmann solvers mentioned above.  
For the sake of simplicity, within this work we will run our simulations in vacuum and do not add a solution contribution since we already including water molecules and Helmholtz planes as a first-order approximation for the voltage-charge response.
Future work should focus on integrating more detailed knowledge of the surrounding solution into the voltage drop within the solution.

To relate the calculated work function from DFT to the equilibrium voltage of a DFT calculation, we will use the Trasatti convention,\cite{trasatti1986absolute} where we define the potential of a given DFT slab as (versus SHE) as 
\begin{equation}
    \Phi = W/|e| - 4.44\text{ V},
\end{equation}
where $W$ is the work function of the system, defined as the opposite of the DFT calculated Fermi level, i.e. $W=-\varepsilon_{\text{F}}$.
(This relation is only strictly true for pure water, but we will accept the approximation for this system).
Thus, by using this methodology, we can specify a charge $Q$ on the electrode, and then test different distributions of $q_{\text{h}}^{\text{ox.}|\text{sol.}}$ and  $q_{\text{DFT}}^{\text{ox.}|\text{sol.}}$ to find the equilibrium charge state and voltage of the system.
This allows standard DFT slab calculations of metal-insulator-solution interfaces to fully include voltage effects-- without the need to simulate 10s of nm of material.

By reducing the number of atoms needed to simulate within DFT by at least an order of magnitude, this work significantly speeds up calculations of realistic voltages for the system, to the point of making $>$ 5 nm systems computationally feasible.
Nevertheless, several smaller DFT calculations are needed to optimize the free energy within the system, finding the charge distribution which leads to a Fermi level being in equilibrium throughout the interface. 
In this work, we manually test different combinations of charge to find the correct equilibrium distribution, which leads to roughly five to ten DFT slab calculations needed for each data point.
We have  previously implemented an automated Newton-Raphson optimization algorithm for finding equilibrium free energy within semiconductor-solution systems.\cite{campbell2019voltage}
A similar algorithm could also be extended to these metal-insulator-solution systems for automated voltage processing.

\subsection*{DFT Details}

\begin{figure*}
    \centering
    \includegraphics[width=\textwidth]{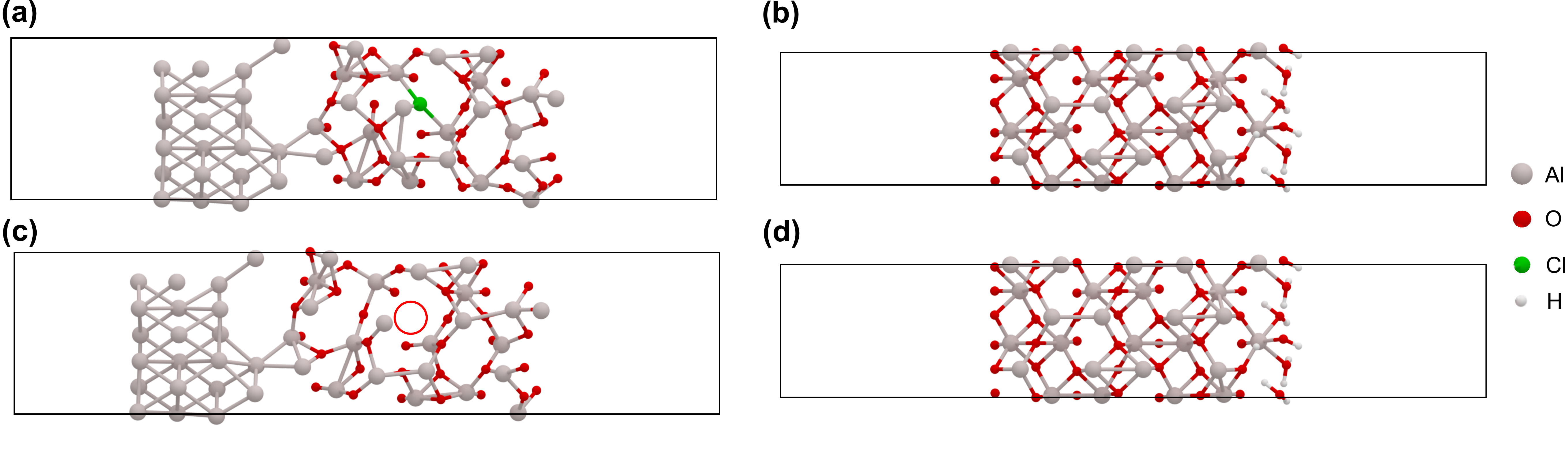}
    \caption{The structures used for the Al(0001)/$\alpha$-Al$_2$O$_3$-water interface calculations with (a) the Al(0001)/$\alpha$-Al$_2$O$_3$ interface with Cl inserted, (b) the $\alpha$-Al$_2$O$_3$/water interface used for the Cl calculations, (c) the Al(0001)/$\alpha$-Al$_2$O$_3$ interface with an oxygen vacancy, highlighted by the red circle, and (d) the $\alpha$-Al$_2$O$_3$/water interface used for the O-vacancy calculations.}
    \label{fig:no-gb}
\end{figure*}

\begin{figure*}
    \centering
    \includegraphics[width=\textwidth]{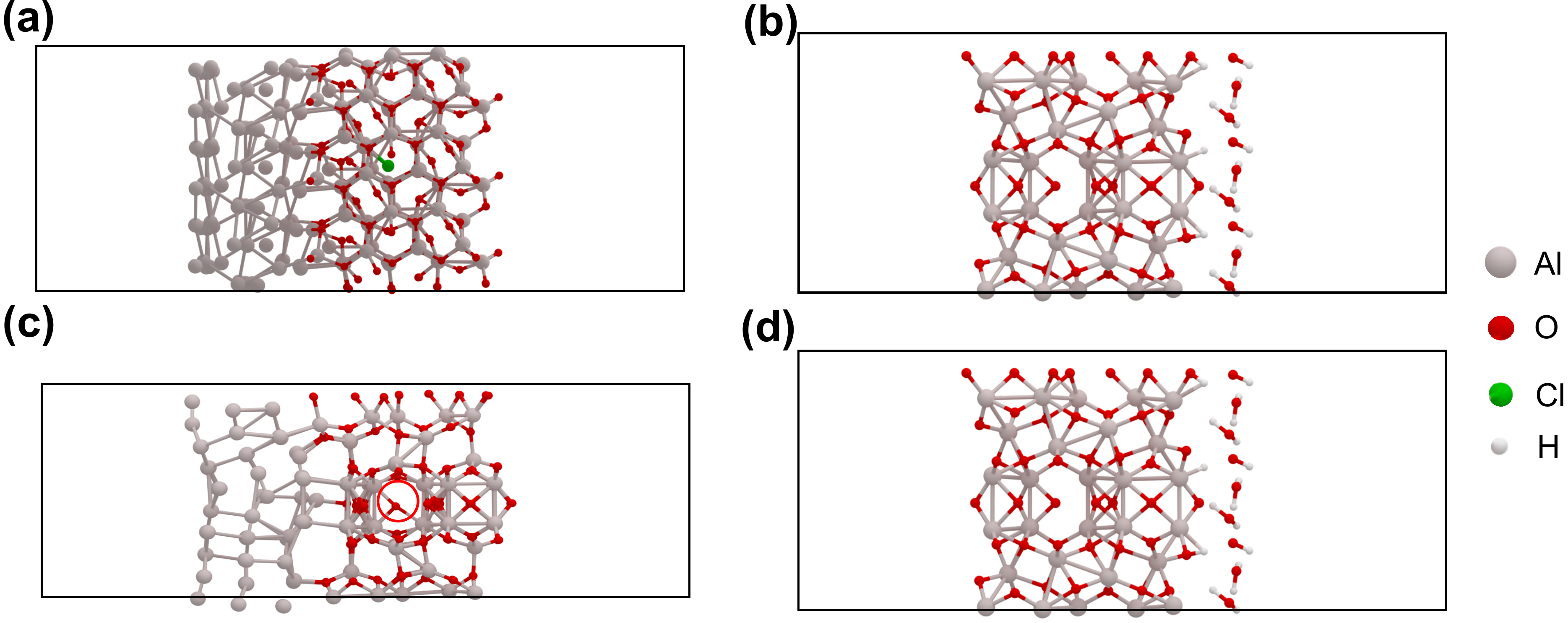}
    \caption{The structures used for the Al(0001)/$\alpha$-Al$_2$O$_3$-water interface calculations including a grain boundary for (a) the Al(0001)/$\alpha$-Al$_2$O$_3$ interface with Cl inserted, (b) the $\alpha$-Al$_2$O$_3$/water interface used for the Cl calculations, (c) the Al(0001)/$\alpha$-Al$_2$O$_3$ interface with an oxygen vacancy, highlighted by the red circle, and (d) the $\alpha$-Al$_2$O$_3$/water interface used for the O-vacancy calculations.}
    \label{fig:gb}
\end{figure*}

We use structures based on the interfaces used in Leung's work on Al surfaces.\cite{leung2021first}
Leung investigated chloride insertion into Al(111)/$\alpha$-Al$_2$O$_3$(0001) surfaces, finding that chloride insertion was only favorable in situations with grain boundaries.
By using these atomic structures, we can make more direct comparisons with the Leung work.
The $\alpha$ phase of Al$_2$O$_3$ is selected as an approximation of the amorphous alumina that would develop on a surface in contact with water molecules. 
The Al/$\alpha$-Al$_2$O$_3$ structure is based on the ``FCC'' interface structure,\cite{siegel2002adhesion} but with double lattice cell dimensions.
We work with two interfaces throughout the work, one without a grain boundary, shown in Fig.~\ref{fig:no-gb}, and one with a grain boundary, shown in Fig.~\ref{fig:gb}. 
For each interface, we create a structure both at the metal-oxide interface and the oxide-solution interface. 
All the oxide-solution structures in question include a monolayer of water on the surface representing a buildup from humidity.
In Fig.~\ref{fig:no-gb}b and d, this represents 6 water molecules, and in Fig.~\ref{fig:gb}b and d this represents 12 water molecules. 
The arrangement of water molecules in this work was taken from Leung's work,\cite{leung2021first} which was from a snapshot from \textit{ab-initio} molecular dynamics (AIMD) chosen to be typical of a system at equilibrium.
The exact arrangement of water molecules has been shown to have significant effects on the total work function of DFT calculations.\cite{cheng2012alignment,kharche2014first,hormann2019absolute}
At non-zero temperatures, the water molecules will necessarily be dynamic, and this represents a reasonable snapshot.
By keeping the arrangement of water molecules consistent between the structures with and without a grain boundary, moreover, we minimize the impacts this has on the relative error between the two structures. 
The structures used in this work have been included in the supplementary files. 

All electronic structure calculations are done using the {\sc quantum espresso} package~\cite{giannozzi2009quantum}.
We use norm-conserving pseudopotentials from the PseudoDojo repository~\cite{van2018pseudodojo} and the Perdew-Burke-Ernzerhof exchange-correlation functional.\cite{perdew1996generalized}
The Al atom uses a Neon core with a valence configuration of 3s$^2$3p$^1$ (Z=3), the O atom uses a Helium core with a 2s$^2$2p$^4$ valence configuration (Z=6), and the Cl atom uses a Neon core with a valence configuration of 3s$^2$3p$^5$ (Z=7).
We use kinetic energy cutoffs of 50 Ry and 200 Ry for the plane wave basis sets used to describe the Kohn-Sham orbitals and charge density, respectively.
We use a 2$\times$2$\times$1 Monkhorst-Pack grid~\cite{monkhorst1976special} to sample the Brillioun zone in our calculations.
Ionic relaxations are done until forces are converged below 50 meV/\AA\ and the electronic calculations are done with a 10$^{-10}$ Ry energy threshold.
All atoms were relaxed and none frozen. 

We use the {\sc Environ} package \cite{andreussi2012revised} to calculate the parabolic corrections to the surface dipole.
For simplicity, all calculations are done in vacuum, with a vacuum region of $\geq$ 20 \AA.

In the results calculated from this work, we use the following parameterizations for the QCA equations outlined above: $\epsilon_{\text{sc}} = 9.1$ to match Al$_2$O$_3$, $p_d = 0$, $n_d = 10^{15}$ cm$^{-3}$, $T = 300$ K, and $c^{\circ} = 0 $ since we are simplifying the problem as in vacuum and thus not including Poisson-Boltzmann contributions from the solution. 
We instead use the Helmholtz distribution of ionic charge in the solution region combined with an explicit water monolayer, which his often used as a first approximation of full ionic distributions.\cite{schmickler2010interfacial}

\section*{Results}

We first consider chloride ion insertion in an interface without any grain boundaries. 
To start, we calculate the flatband potential of the system before any oxygen vacancies or chlorine atoms are inserted. 
For the oxide-solution system we find a Fermi level of $ \varepsilon_{\text{F,fb}}^{\text{ox.}|\text{sol.}} = -6.2$ eV and for the metal-oxide system we find $\varepsilon_{\text{F,fb}}^{\text{m.}|\text{ox.}} = -6.1$ eV. 
To find the flatband potential of the total system it is necessary to find the arrangement of charge that causes the Fermi level to be in equilibrium throughout the system, and the total charge on the electrode must be neutral, i.e. $Q=0$ .
This constrains the problem sufficiently that the flatband potential of the entire system can be calculated as $\Phi_{\text{fb}} \approx 1.7 $ V (SHE).

We then investigate placing an oxygen vacancy into the structure. 
Following Leung's work, we place the oxygen vacancy 4.3 \AA\ away from the metal surface, where the zero $z$-location is here defined as the first oxygen atoms of the oxide structure moving away from the metal.
We allow the system to relax around the vacancy, which is uncharged. 
We then place a chlorine atom into this oxygen vacancy and again allow the system to relax. 
We assume that oxygen vacancies are not mobile within the timeframe of chloride insertion. 
For the sake of this demonstration, we do not examine the kinetics of chloride insertion, but the QCA techniques presented here do provide an opportunity for future work calculating kinetic barriers while the surface is under applied voltage. 
When we double the lateral size of the system, we observe a -0.16 V in the calculated flatband potential for the insertion of the Cl ion, indicating that the size of the system is reasonably converged so that concentration effects will not overwhelm the calculation. 
We examine a number of changes to the DFT system and their impact on the Cl insertion flatband potential in the Appendix. 
Using Bader charge analysis,\cite{henkelman2006fast} we confirm that the Cl atom carries $\approx$ 8 electrons, and thus is, in fact, acting as a Cl$^{-}$ ion within the system. 

\begin{figure}
    \centering
    \includegraphics[width=0.6\textwidth]{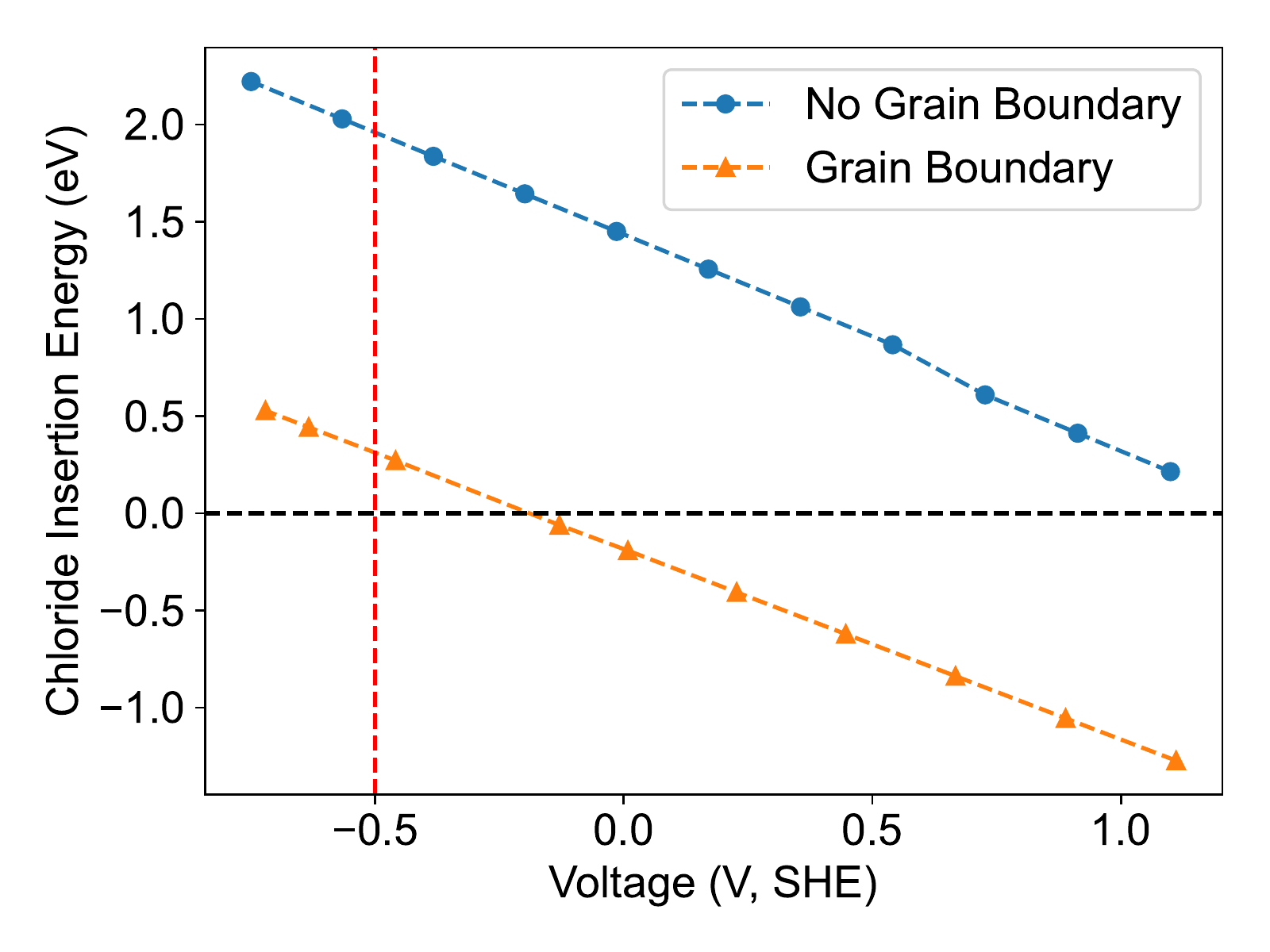}
    \caption{A stability diagram indicating the calculated insertion energy of a chloride ion as a function of voltage.
    When the chloride ion is inserted into a grain boundary it becomes stable (i.e. $< 0$ eV, indicated by the horizontal black dashed line) at voltages $> -0.2$ V (SHE).
    This compares favorably with the experimentally measured onset of chloride pitting at -0.5 V (SHE) indicated by the vertical red dashed line. 
    In contrast, in systems without grain boundaries, chloride ion insertion is not favorable until much higher voltages. }
    \label{fig:stability-diagram}
\end{figure}

We then apply QCA as outlined in above over a variety of voltages, and calculate the resulting formation energy for chloride insertion into the oxygen vacancy. 
Here, the DFT energies for the chloride insertion and oxygen vacancy, $E_{\text{DFT}}(\text{Cl ins.})$ and $E_{\text{DFT}}(\text{O vac.})$ in Eq.~\ref{eq:total-e} need to be calculated including the impact of voltage as found in the equilibrium QCA configuration. 
As seen in Fig.~\ref{fig:stability-diagram}, the formation energy for chloride insertion is highly positive across a wide range of voltages, implying that chloride is highly unlikely to form at these potentials. 
The chloride insertion energy only dips below 0 eV at $\gtrapprox$ 1.2 V (SHE). 
This matches previous work by Leung who reports that chloride ions are not stable inserted into crystalline $\alpha$-Al$_2$O$_3$.\cite{leung2021first}

We predict that the relationship between the applied voltage and the chloride insertion energy is largely linear.
This indicates that the dominant term in the energy is in fact the chloride ion energy introduced in Eqn.~\ref{eq:chloride-e}.
This can be attributed to a majority of the charge in the system accumulating at the oxide-solution interface, leaving only a small potential drop in the bulk oxide region, where the parabolic relationships for voltage would become more dominant. 
This also highlights the importance of pairing QCA with a computational standard hydrogen electrode or similar approach which allows for inclusion of the voltage in the calculation of the ionic energy.

We next consider an interface that includes grain boundaries. 
We again first measure the Fermi level of the oxide-solution system as $ \varepsilon_{\text{F,fb}}^{\text{ox.}|\text{sol.}} = -5.54$ eV and the metal-oxide system as $\varepsilon_{\text{F,fb}}^{\text{m.}|\text{ox.}} = -6.1$ eV.
Using the same procedure as before we can thus find a flatband potential for the grain boundary system of $\Phi_{\text{fb}} \approx 1.4 $ V (SHE).
We then insert oxygen vacancies or chlorine atoms into the structure and allow the system to relax. 
It should be noted that due to periodic boundary conditions, the density of grain boundaries is much higher in this system than would be realistically expected. 
We again confirm with Bader charge analysis that the Cl atom is acting as a Cl$^{-}$ ion within the system. 

As seen in Fig.~\ref{fig:stability-diagram}, we then test a variety of voltages across the system with grain boundaries. 
We predict that chloride ion insertion becomes stable at voltages $> -0.2$ V (SHE).
This correlates relatively well with the previous experiments showing pitting corrosion at voltages $> -0.5$ V (SHE)  \cite{dibari1971electrochemical,bessone1992eis} (see below for a discussion on comparing theoretical and experimental voltages).
It also matches previous results from Leung, who similarly finds that chloride ion insertion is only favorable at grain boundaries within the aluminum oxide.\cite{leung2021first} 
This work emphasizes the role that grain boundaries likely play in pitting corrosion according to the point defect model, serving as locations for chloride insertion into the aluminum oxide and serving as corrosion centers.
This result also follows intuition: Cl$^-$ is a larger anion than oxygen and since the grain boundary is less dense than the surrounding Al$_2$O$_3$ inserting the Cl in this grain boundary causes less distortion to the underlying lattice and is thus more favorable.

\begin{figure}
    \centering
    \includegraphics[width=0.6\textwidth]{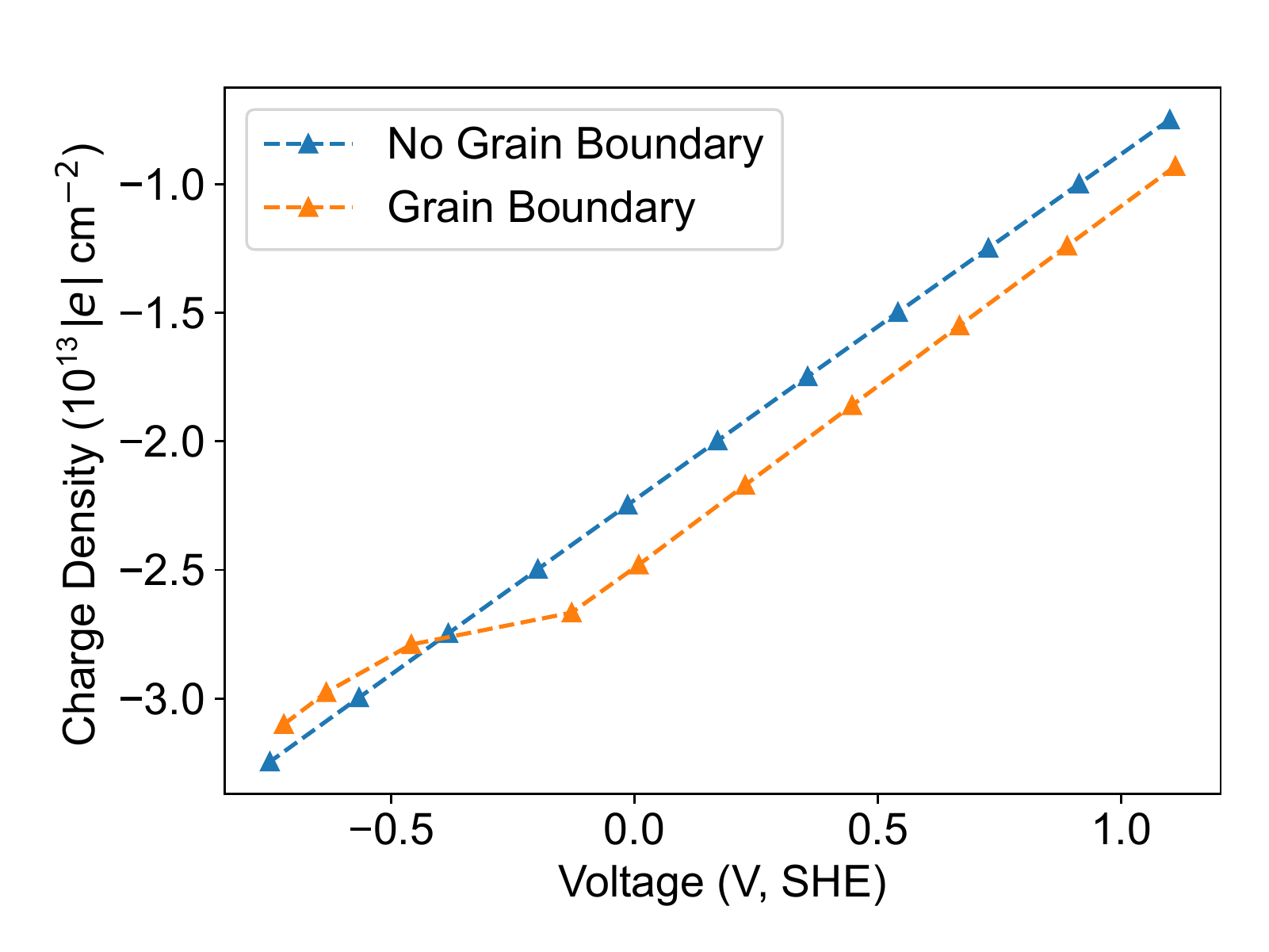}
    \caption{The equilibrium charge density at the interface with a Cl$^{-}$ ion inserted both with and without grain boundaries. 
    Here, negative values are taken to indicate excessive electrons. }
    \label{fig:chg-density}
\end{figure}

The voltage dependence of the charge density, shown in Fig.~\ref{fig:chg-density}, also matches the intuitive picture of chloride stability in grain boundaries we have established.
For interfaces with and without a grain boundary, the charge density is largely linear as a function of voltage. 
This reflects the fact that, as previously noted, the majority of the charge in our simulations of the system is caught at the oxide/solution interface, causing the interface to behave in a more metallic manner, particularly away from the flatband potential.
Furthermore, the interface with grain boundaries consistently has lower charge densities at voltages where the Cl$^{-}$ ion is predicted to be stable, somewhat matching the intuition that more charge is allowed in the system. 
(In this notation, ``lower'' charge densities indicate a greater density of electrons induced at the interface.)
This is only violated at $\Phi \lessapprox$ -0.2 V (SHE), corresponding to the location where a Cl$^{-}$ ion is no longer stable in the system.
Below this voltage, the charge density for the two systems becomes nearly equal.

\section*{Discussion}
\label{sec:discussion}
While QCA provides a useful tool for attaching relative voltages to DFT calculations of interfaces, it does not solve a variety of outstanding problems in defining the electronic voltage of a neutral DFT system relative to experiment. 
In particular, the arrangement of water molecules within the first monolayer of coverage has been shown to have a large impact on the estimated flatband potential, with previous studies on this system showing the impact of different H$_2$O configurations being 0.2-0.5 eV.\cite{sakong2016structure,sakong2018electric} 
This is a difficult problem to overcome as there is no clearly ``correct'' configuration of the water molecules on the surface.
One possible solution would be to average over several possible H$_2$O configurations sampled from an AIMD calculation and attempt to form some averaging.

As indicated previously, another concern is the relative concentration of charged defects being examined. 
The periodic nature of these DFT calculations ensures that the 2D density of any defects will be relatively high.
Furthermore, realistic systems likely have larger macrostructure, with grain boundaries and defects that are not actually periodic in nature. 
This can be partially overcome by creating larger lateral system sizes and performing convergence tests.
Increasing the lateral size of the interfaces typically necessitates a a tradeoff of the length of the slab to keep calculations computationally feasible. 
This shorter slab can still be accommodated to provide useful results with QCA provided that a few layers have been included around the defect and interface. 

These twin uncertainties make exact comparison of our predicted chloride insertion onset voltage to macroscopic potentials extracted from experiment problematic. 
Our QCA predictions of exact potentials should then be taken as semi-quantitative about the system, not exact.
For instance, the finding that grain boundaries within the aluminum oxide make chloride insertion much more favorable are consistent regardless of the exact voltage at which they become favorable. 
Similarly, if our results predicted that chloride insertion was only favorable only at voltages $>$ 3 V (SHE), it would be likely that we were not correctly representing the interface within DFT (or that QCA itself was an incorrect framework for the problem).
When predicting relative voltage differences in the same system (i.e. the difference between two voltages predicted using the same surface termination), however, QCA should be more rigorous as the previous drawbacks to exact work function determination are held constant.

QCA nevertheless presents an advantage over other computational approaches to controlling the voltage of an electrode by coupling DFT to continuum and Poisson-Boltzmann descriptions of the solution. 
This is namely the ability to include an extended insulating (oxide) region within the simulation.
Previous work has needed to focus on metals with a linear relationship between the charge and voltage of the system, consistent with a constant capacitance model.\cite{schnur2011challenges,keilbart2017quantum,haruyama2018electrode,kano2021study}
While these approaches can be quite useful, their utility can be limited in comparison to experiment, especially if large passivating layers are expected to form over the metal electrode and the capacitance can no longer be expected to be constant as a function of potential. 
Given that the contact layer between metal and oxide has been shown to sometimes have barriers as high as 2-3 V,\cite{yu2019combining,cornette2020relation} the impact of the extended oxide layer is vital to fully understanding the voltage behavior of an electrode. 
QCA thus presents a computationally scalable method for addressing voltages in systems with extended insulating regions.

\section*{Conclusion and Outlook}
In this work, we have extended the QCA methodology to encompass metal-oxide interfaces that are also in contact with a solution. 
This extension allows us to provide first principles calculations of atomic mechanisms behind corrosion and other electrochemical processes as a function of applied voltage, which has previously been difficult to manage within a DFT framework. 
We demonstrate this methodology on an Al/$\alpha$-Al$_2$O$_3$ interface, calculating that chloride insertion is only favorable above -0.2 V (SHE) when a grain boundary is included at the interface.
This conclusion is broadly consistent with both previous first principles work, and experimental measurements.
This provides helpful context for parameterizing the point defect model with first-principles results.

While QCA provides a useful method for examining interfacial problems that includes voltage, it does not resolve some of the traditional difficulties of DFT.
Band gaps will be underestimated in semiconductors and insulators when using semi-local functionals.
As long as the potential drop across the electrode is less than the band gap of the DFT material, however, this should not affect the relative voltage results. 
Similarly, the exact atomic structures of both the metal-oxide and oxide-solution interfaces exert a large influence on the assigned flatband potential of the interfacial structure. 
Numerous DFT methods have been developed to address this issue, with high temperature ab-initio molecular dynamics providing one option for generating structures.
Larger surveys of possible structures should still be made for QCA analysis of any given system. 
It nevertheless presents advantages over previous work in being able to examine systems with large passivation layers. 

In this work, we demonstrated QCA in the context of corrosion of aluminum, but it can clearly also be used in numerous applications of interest.
In battery electrode research, for instance, the exact atomistic interaction of a solid-electrolyte interface (SEI) with Li ions shuttling through the structure is often not well understood as a function of voltage.\cite{leung2020dft} 
QCA provides one potential route for exploring these systems while including a realistic voltage to the system.
Another application may be photocatalysis, where a semiconductor is placed in contact with solution to catalyze a reaction, often attached to a metal electrode to provide some level of voltage to the system.\cite{wenderich2016methods}
Beyond electrochemistry, a similar methodology can be applied to electronic interfaces of interest such as p-n junctions and Schottky barriers.\cite{subramanian2020photophysics} 
Thus, QCA provides a new tool for computational electrochemistry across a broad variety of interfaces and applications, including voltage effects at reasonable computational cost. 

\ack
We would like to thank Kevin Leung for providing the initial Al/Al$_2$O$_3$ system and a critical reading of the draft manuscript, as well as Andrew Baczewski for useful feedback.
This work was supported by the Laboratory Directed Research and Development program at Sandia National Laboratories under project 226354 and 229741.
Sandia National Laboratories is a multi-mission laboratory managed and operated by National Technology and Engineering Solutions of Sandia, LLC, a wholly owned subsidiary of Honeywell International, Inc., for DOE's National Nuclear Security Administration under contract DE-NA0003525.
This paper describes objective technical results and analysis.
Any subjective views or opinions that might be expressed in the paper do not necessarily represent the views of the U.S. Department of Energy or the United States Government.
\appendix
\section{Dependence of the flatband voltage of the Cl$^{-}$ insertion on DFT system}
\setcounter{section}{1}

The results presented in the main text are heavily dependent on the flatband potential calculated for Cl$^{-}$ insertion. 
It is thus natural to investigate how this flatband potential is dependent on the exact configuration used for our DFT calculations. 
We investigate a number of alternatives in the no grain boundary case, such as doubling the lateral area of the system, moving the position of the Cl insertion, and increasing the number of oxide layers modeled, with the results displayed in Table~\ref{tab:extra}. 
We find that while the exact number changes, the flatband voltage is largely robust to these changes, with the exception of moving the Cl ion closer or farther from the metal-oxide interface.
This matches the previous work from Leung.\cite{leung2021first}
We additionally tested increasing the number of oxide layers and found a change of -0.01 V in the flatband potential.
These results highlight the importance of converging system size with respect to flatband potential. 
Once the system is sufficiently converged, the length of the DFT region included within the QCA framework outlined in the main text should not impact the results.

\begin{table}
\caption{Flatband voltage of different system configurations for Cl insertion}
\begin{center}
\begin{tabular}{ |l|r|c| } 
 \hline
 System & Flatband Voltage (V, SHE) \\ 
 \hline
 Cl inserted cell (Fig. 3a) & 1.66 \\ 
 Lateral doubled Cell & 1.50 \\ 
 Cl$^{-}$ position changed, same distance from m$|$ox. interface & 1.60 \\ 
 Cl$^{-}$ position changed, larger distance from m$|$ox. interface &  0.80 \\ 
 Increase in number of oxide layers modeled &  1.65 \\ 
 \hline
 
\end{tabular}
\end{center}
\label{tab:extra}
\end{table}

\section{Supporting Files}
We have added supporting files which include the scripts used to generate QCA calculations and input structures for both the calculations with and without grain boundaries. 

\section*{References}
\bibliographystyle{unsrt}
\bibliography{refs}

\begin{thebibliography}{10}

\bibitem{koch2002corrosion}
G.~H. Koch, M.~P. Brongers, N.~G. Thompson, Y.~P. Virmani, J.~H. Payer, et~al.
\newblock Corrosion cost and preventive strategies in the United States.
\newblock Technical report, United States. Federal Highway Administration,
  (2002).

\bibitem{dibari1971electrochemical}
G.~Dibari and H.~Read.
\newblock Electrochemical behavior of high purity aluminum in chloride
  containing solutions.
\newblock {\em Corrosion}, 27(11), 483--494, (1971).

\bibitem{bessone1992eis}
J.~Bessone, D.~Salinas, C.~Mayer, M.~Ebert, and W.~Lorenz.
\newblock An EIS study of aluminium barrier-type oxide films formed in
  different media.
\newblock {\em Electrochimica Acta}, 37(12), 2283--2290, (1992).

\bibitem{lin1981point}
L.~Lin, C.~Chao, and D.~Macdonald.
\newblock A point defect model for anodic passive films: II. Chemical breakdown
  and pit initiation.
\newblock {\em Journal of the Electrochemical Society}, 128(6), 1194, (1981).

\bibitem{chao1981point}
C.~Chao, L.~Lin, and D.~Macdonald.
\newblock A point defect model for anodic passive films: I. Film growth
  kinetics.
\newblock {\em Journal of the Electrochemical Society}, 128(6), 1187, (1981).

\bibitem{engelhardt2004unification}
G.~Engelhardt and D.~D. Macdonald.
\newblock Unification of the deterministic and statistical approaches for
  predicting localized corrosion damage. I. Theoretical foundation.
\newblock {\em Corrosion Science}, 46(11), 2755--2780, (2004).

\bibitem{frankel1998pitting}
G.~Frankel.
\newblock Pitting corrosion of metals: a review of the critical factors.
\newblock {\em Journal of the Electrochemical Society}, 145(6), 2186, (1998).

\bibitem{natishan20182017}
P.~Natishan.
\newblock 2017 wr whitney award: Perspectives on chloride interactions with
  passive oxides and oxide film breakdown.
\newblock {\em Corrosion}, 74(3), 263--275, (2018).

\bibitem{evertsson2015thickness}
J.~Evertsson, F.~Bertram, F.~Zhang, L.~Rullik, L.~Merte, M.~Shipilin,
  M.~Soldemo, S.~Ahmadi, N.~Vinogradov, F.~Carl{\`a}, et~al.
\newblock The thickness of native oxides on aluminum alloys and single
  crystals.
\newblock {\em Applied Surface Science}, 349, 826--832, (2015).

\bibitem{brown1973use}
F.~Brown and W.~Mackintosh.
\newblock The use of Rutherford backscattering to study the behavior of
  ion-implanted atoms during anodic oxidation of Aluminum: Ar, Kr, Xe, K, Rb,
  Cs, Cl, Br, and l.
\newblock {\em Journal of The Electrochemical Society}, 120(8), 1096, (1973).

\bibitem{shimizu1999novel}
K.~Shimizu, G.~Brown, K.~Kobayashi, P.~Skeldon, G.~Thompson, and G.~Wood.
\newblock A novel approach for the study of the migration of Cl- ions in anodic
  alumina.
\newblock {\em Corrosion Science}, 41(9), 1835--1847, (1999).

\bibitem{serna2006critical}
L.~Serna, K.~Zavadil, C.~Johnson, F.~Wall, and J.~Barbour.
\newblock A critical implanted cl concentration for pit initiation on aluminum
  thin films.
\newblock {\em Journal of The Electrochemical Society}, 153(8), B289, (2006).

\bibitem{kim2013nature}
E.-G. Kim and J.-L. Br{\'e}das.
\newblock The nature of the aluminum--aluminum oxide interface: A nanoscale
  picture of the interfacial structure and energy-level alignment.
\newblock {\em Organic Electronics}, 14(2), 569--574, (2013).

\bibitem{batyrev2001plane}
I.~G. Batyrev and L.~Kleinman.
\newblock In-plane relaxation of Cu (111) and Al (111)/$\alpha$- Al 2 O 3
  (0001) interfaces.
\newblock {\em Physical Review B}, 64(3), 033410, (2001).

\bibitem{koberidze2018structural}
M.~Koberidze, M.~Puska, and R.~Nieminen.
\newblock Structural details of Al/Al 2 O 3 junctions and their role in the
  formation of electron tunnel barriers.
\newblock {\em Physical Review B}, 97(19), 195406, (2018).

\bibitem{costa2014atomistic}
D.~Costa, T.~Ribeiro, F.~Mercuri, G.~Pacchioni, and P.~Marcus.
\newblock Atomistic modeling of corrosion resistance: a first principles study
  of O2 reduction on the Al (111) surface covered with a thin hydroxylated
  alumina film.
\newblock {\em Advanced Materials Interfaces}, 1(3), 1300072, (2014).

\bibitem{siegel2002adhesion}
D.~J. Siegel, L.~G. Hector~Jr, and J.~B. Adams.
\newblock Adhesion, atomic structure, and bonding at the Al (111)/$\alpha$- Al
  2 O 3 (0001) interface: A first principles study.
\newblock {\em Physical Review B}, 65(8), 085415, (2002).

\bibitem{weber2009point}
J.~Weber, A.~Janotti, and C.~Van~de Walle.
\newblock Point defects in Al2O3 and their impact on gate stacks.
\newblock {\em Microelectronic Engineering}, 86(7-9), 1756--1759, (2009).

\bibitem{carrasco2004theoretical}
J.~Carrasco, J.~R. Gomes, and F.~Illas.
\newblock Theoretical study of bulk and surface oxygen and aluminum vacancies
  in $\alpha$- Al 2 O 3.
\newblock {\em Physical Review B}, 69(6), 064116, (2004).

\bibitem{hine2009supercell}
N.~Hine, K.~Frensch, W.~Foulkes, and M.~Finnis.
\newblock Supercell size scaling of density functional theory formation
  energies of charged defects.
\newblock {\em Physical Review B}, 79(2), 024112, (2009).

\bibitem{janetzko2004first}
F.~Janetzko, R.~A. Evarestov, T.~Bredow, and K.~Jug.
\newblock First-principles periodic and semiempirical cyclic cluster
  calculations for single oxygen vacancies in crystalline Al2O3.
\newblock {\em physica status solidi (b)}, 241(5), 1032--1040, (2004).

\bibitem{liu2019dft}
M.~Liu, Y.~Jin, C.~Leygraf, and J.~Pan.
\newblock A DFT-Study of Cl Ingress into $\alpha$-Al2O3 (0001) and Al (111) and
  its possible influence on localized corrosion of Al.
\newblock {\em Journal of the Electrochemical Society}, 166(11), C3124, (2019).

\bibitem{liu2021density}
M.~Liu, Y.~Jin, B.~Chen, C.~Leygraf, L.~Wang, and J.~Pan.
\newblock Density Functional Theory Study of Influence of Oxide Thickness and
  Surface Alloying on Cl Migration within $\alpha$-Al2O3.
\newblock {\em Journal of The Electrochemical Society}, 168(8), 081508, (2021).

\bibitem{leung2021first}
K.~Leung.
\newblock First principles, explicit interface studies of oxygen vacancy and
  chloride in alumina films for corrosion applications.
\newblock {\em Journal of The Electrochemical Society}, 168(3), 031511, (2021).

\bibitem{norskov2004origin}
J.~K. N{\o}rskov, J.~Rossmeisl, A.~Logadottir, L.~Lindqvist, J.~R. Kitchin,
  T.~Bligaard, and H.~Jonsson.
\newblock Origin of the overpotential for oxygen reduction at a fuel-cell
  cathode.
\newblock {\em The Journal of Physical Chemistry B}, 108(46), 17886--17892,
  (2004).

\bibitem{rossmeisl2007electrolysis}
J.~Rossmeisl, Z.-W. Qu, H.~Zhu, G.-J. Kroes, and J.~K. N{\o}rskov.
\newblock Electrolysis of water on oxide surfaces.
\newblock {\em Journal of Electroanalytical Chemistry}, 607(1-2), 83--89,
  (2007).

\bibitem{man2011universality}
I.~C. Man, H.-Y. Su, F.~Calle-Vallejo, H.~A. Hansen, J.~I. Mart{\'\i}nez, N.~G.
  Inoglu, J.~Kitchin, T.~F. Jaramillo, J.~K. N{\o}rskov, and J.~Rossmeisl.
\newblock Universality in oxygen evolution electrocatalysis on oxide surfaces.
\newblock {\em ChemCatChem}, 3(7), 1159--1165, (2011).

\bibitem{atkins2006physical}
P.~Atkins and J.~De~Paula.
\newblock {\em Physical Chemistry}, volume~8.
\newblock W. H. Freeman and Company, (2006).

\bibitem{le1971dissociation}
R.~J. Le~Roy and R.~B. Bernstein.
\newblock Dissociation energies and long-range potentials of diatomic molecules
  from vibrational spacings: The halogens.
\newblock {\em Journal of Molecular Spectroscopy}, 37(1), 109--130, (1971).

\bibitem{campbell2017quantum}
Q.~Campbell and I.~Dabo.
\newblock Quantum-continuum calculation of the surface states and electrical
  response of silicon in solution.
\newblock {\em Physical Review B}, 95(20), 205308, (2017).

\bibitem{campbell2019voltage}
Q.~Campbell, D.~Fisher, and I.~Dabo.
\newblock Voltage-dependent reconstruction of layered Bi 2 WO 6 and Bi 2 MoO 6
  photocatalysts and its influence on charge separation for water splitting.
\newblock {\em Physical Review Materials}, 3(1), 015404, (2019).

\bibitem{subramanian2020photophysics}
S.~Subramanian, Q.~T. Campbell, S.~K. Moser, J.~Kiemle, P.~Zimmermann,
  P.~Seifert, F.~Sigger, D.~Sharma, H.~Al-Sadeg, M.~Labella~III, et~al.
\newblock Photophysics and electronic structure of lateral graphene/MoS2 and
  metal/MoS2 junctions.
\newblock {\em ACS Nano}, 14(12), 16663--16671, (2020).

\bibitem{schmickler2010interfacial}
W.~Schmickler and E.~Santos.
\newblock {\em Interfacial Electrochemistry}.
\newblock Springer Science \& Business Media, (2010).

\bibitem{bardeen1947surface}
J.~Bardeen.
\newblock Surface states and rectification at a metal semi-conductor contact.
\newblock {\em Physical Review}, 71(10), 717, (1947).

\bibitem{cowley1965surface}
A.~Cowley and S.~Sze.
\newblock Surface states and barrier height of metal-semiconductor systems.
\newblock {\em Journal of Applied Physics}, 36(10), 3212--3220, (1965).

\bibitem{bard1980role}
A.~J. Bard, F.-R.~F. Fan, A.~S. Gioda, G.~Nagasubramanian, and H.~S. White.
\newblock On the role of surface states in semiconductor electrode
  photoelectrochemical cells.
\newblock {\em Faraday Discussions of the Chemical Society}, 70, 19--31,
  (1980).

\bibitem{keilbart2017quantum}
N.~Keilbart, Y.~Okada, A.~Feehan, S.~Higai, and I.~Dabo.
\newblock Quantum-continuum simulation of the electrochemical response of
  pseudocapacitor electrodes under realistic conditions.
\newblock {\em Physical Review B}, 95(11), 115423, (2017).

\bibitem{fisicaro2016generalized}
G.~Fisicaro, L.~Genovese, O.~Andreussi, N.~Marzari, and S.~Goedecker.
\newblock A generalized Poisson and Poisson-Boltzmann solver for electrostatic
  environments.
\newblock {\em The Journal of Chemical Physics}, 144(1), 014103, (2016).

\bibitem{mathew2019implicit}
K.~Mathew, V.~C. Kolluru, S.~Mula, S.~N. Steinmann, and R.~G. Hennig.
\newblock Implicit self-consistent electrolyte model in plane-wave
  density-functional theory.
\newblock {\em The Journal of Chemical Physics}, 151(23), 234101, (2019).

\bibitem{trasatti1986absolute}
S.~Trasatti.
\newblock The absolute electrode potential: an explanatory note
  (Recommendations 1986).
\newblock {\em Pure and Applied Chemistry}, 58(7), 955--966, (1986).

\bibitem{cheng2012alignment}
J.~Cheng and M.~Sprik.
\newblock Alignment of electronic energy levels at electrochemical interfaces.
\newblock {\em Physical Chemistry Chemical Physics}, 14(32), 11245--11267,
  (2012).

\bibitem{kharche2014first}
N.~Kharche, J.~T. Muckerman, and M.~S. Hybertsen.
\newblock First-principles approach to calculating energy level alignment at
  aqueous semiconductor interfaces.
\newblock {\em Physical Review Letters}, 113(17), 176802, (2014).

\bibitem{hormann2019absolute}
N.~G. H{\"o}rmann, Z.~Guo, F.~Ambrosio, O.~Andreussi, A.~Pasquarello, and
  N.~Marzari.
\newblock Absolute band alignment at semiconductor-water interfaces using
  explicit and implicit descriptions for liquid water.
\newblock {\em npj Computational Materials}, 5(1), 1--6, (2019).

\bibitem{giannozzi2009quantum}
P.~Giannozzi, S.~Baroni, N.~Bonini, M.~Calandra, R.~Car, C.~Cavazzoni,
  D.~Ceresoli, G.~L. Chiarotti, M.~Cococcioni, I.~Dabo, et~al.
\newblock QUANTUM ESPRESSO: a modular and open-source software project for
  quantum simulations of materials.
\newblock {\em Journal of Physics: Condensed Matter}, 21(39), 395502, (2009).

\bibitem{van2018pseudodojo}
M.~J. van Setten, M.~Giantomassi, E.~Bousquet, M.~J. Verstraete, D.~R. Hamann,
  X.~Gonze, and G.-M. Rignanese.
\newblock The PseudoDojo: Training and grading a 85 element optimized
  norm-conserving pseudopotential table.
\newblock {\em Computer Physics Communications}, 226, 39--54, (2018).

\bibitem{perdew1996generalized}
J.~P. Perdew, K.~Burke, and M.~Ernzerhof.
\newblock Generalized gradient approximation made simple.
\newblock {\em Physical Review Letters}, 77(18), 3865, (1996).

\bibitem{monkhorst1976special}
H.~J. Monkhorst and J.~D. Pack.
\newblock Special points for Brillouin-zone integrations.
\newblock {\em Physical Review B}, 13(12), 5188, (1976).

\bibitem{andreussi2012revised}
O.~Andreussi, I.~Dabo, and N.~Marzari.
\newblock Revised self-consistent continuum solvation in electronic-structure
  calculations.
\newblock {\em The Journal of Chemical Physics}, 136(6), 064102, (2012).

\bibitem{henkelman2006fast}
G.~Henkelman, A.~Arnaldsson, and H.~J{\'o}nsson.
\newblock A fast and robust algorithm for Bader decomposition of charge
  density.
\newblock {\em Computational Materials Science}, 36(3), 354--360, (2006).

\bibitem{sakong2016structure}
S.~Sakong, K.~Forster-Tonigold, and A.~Gro{\ss}.
\newblock The structure of water at a Pt (111) electrode and the potential of
  zero charge studied from first principles.
\newblock {\em The Journal of Chemical Physics}, 144(19), 194701, (2016).

\bibitem{sakong2018electric}
S.~Sakong and A.~Gro{\ss}.
\newblock The electric double layer at metal-water interfaces revisited based
  on a charge polarization scheme.
\newblock {\em The Journal of Chemical Physics}, 149(8), 084705, (2018).

\bibitem{schnur2011challenges}
S.~Schnur and A.~Gro{\ss}.
\newblock Challenges in the first-principles description of reactions in
  electrocatalysis.
\newblock {\em Catalysis Today}, 165(1), 129--137, (2011).

\bibitem{haruyama2018electrode}
J.~Haruyama, T.~Ikeshoji, and M.~Otani.
\newblock Electrode potential from density functional theory calculations
  combined with implicit solvation theory.
\newblock {\em Physical Review Materials}, 2(9), 095801, (2018).

\bibitem{kano2021study}
K.~Kano, S.~Hagiwara, T.~Igarashi, and M.~Otani.
\newblock Study on the free corrosion potential at an interface between an Al
  electrode and an acidic aqueous NaCl solution through density functional
  theory combined with the reference interaction site model.
\newblock {\em Electrochimica Acta}, 377, 138121, (2021).

\bibitem{yu2019combining}
X.-x. Yu and L.~D. Marks.
\newblock Combining the physics of metal/oxide heterostructure, interface
  dipole, band bending, crystallography, and surface state to understand
  heterogeneity contrast in oxidation and corrosion.
\newblock {\em Corrosion}, 75(2), 152--166, (2019).

\bibitem{cornette2020relation}
P.~Cornette, D.~Costa, and P.~Marcus.
\newblock Relation between Surface Composition and Electronic Properties of
  Native Oxide Films on an Aluminium-Copper Alloy Studied by DFT.
\newblock {\em Journal of The Electrochemical Society}, 167(16), 161501,
  (2020).

\bibitem{leung2020dft}
K.~Leung.
\newblock DFT modelling of explicit solid--solid interfaces in batteries:
  Methods and challenges.
\newblock {\em Physical Chemistry Chemical Physics}, 22(19), 10412--10425,
  (2020).

\bibitem{wenderich2016methods}
K.~Wenderich and G.~Mul.
\newblock Methods, mechanism, and applications of photodeposition in
  photocatalysis: a review.
\newblock {\em Chemical Reviews}, 116(23), 14587--14619, (2016).

\end{thebibliography}

\end{document}